\documentclass[11pt, a4paper]{article}
\usepackage{amsmath}

\newtheorem{theorem}{Theorem}

\begin{document}

\title{The characterization of two-component (2+1)-dimensional 
integrable  systems of hydrodynamic type}
\author{E.V. Ferapontov and K.R. Khusnutdinova
    \footnote{On the leave from: Institute of Mechanics, Ufa Branch of
the Russian Academy of
    Sciences, Karl Marx Str. 6, Ufa, 450000, Russia.}}
    \date{}
    \maketitle
    \vspace{-7mm}
\begin{center}
Department of Mathematical Sciences \\ Loughborough University \\
Loughborough, Leicestershire LE11 3TU \\ United Kingdom \\[2ex]
e-mails: \\[1ex] \texttt{E.V.Ferapontov@lboro.ac.uk}\\
\texttt{K.Khusnutdinova@lboro.ac.uk}
\end{center}

\bigskip

\begin{abstract}
We obtain the necessary and sufficient conditions for a two-component 
$(2+1)$-dimensional system of hydrodynamic type to possess infinitely 
many hydrodynamic reductions. These conditions are
in involution,  implying that the systems in question are locally 
parametrized by 15 arbitrary constants. It is proved that all such 
systems possess three conservation laws of hydrodynamic type and, 
therefore, are symmetrizable in Godunov's sense. Moreover, all such 
systems are proved to possess a scalar pseudopotential  which plays 
the role of
the `dispersionless Lax pair'. We demonstrate that the class of 
two-component systems possessing a scalar pseudopotential is in fact 
{\it identical}  with the class of systems possessing infinitely many 
hydrodynamic reductions, thus establishing the equivalence of  the two 
possible definitions of the integrability. Explicit linearly 
degenerate  examples are constructed.

\bigskip


\bigskip

\end{abstract}

\newpage

\section{Introduction}

We consider the problem of the classification of
$(2+1)$-dimensional integrable quasilinear systems
\begin{equation}
{\bf u}_t+A({\bf u}) {\bf u}_x+B({\bf u}) {\bf u}_y=0
\label{1}
\end{equation}
where $t, x, y$ are independent variables,  ${\bf u}$ is  an
$m$-component column vector and $A({\bf u}), B({\bf u})$ are $m\times
m$ matrices.  We assume that the system is strictly hyperbolic, that 
is, the generic matrix  of the linear family $\lambda I_m+\mu A + B$ 
has $m$  distinct real eigenvalues.
Following our recent paper \cite{Fer3},  we call the system (\ref{1}) 
{\it integrable} if it possesses `sufficiently many'    exact 
solutions of the form
${\bf u}={\bf u}(R^1, ..., R^n)$ where the {\it Riemann invariants}
$R^1, ..., R^n$ solve a pair of
commuting diagonal systems
\begin{equation}
R^i_t=\lambda^i(R)\ R^i_y, ~~~~ R^i_x=\mu^i(R)\ R^i_y;
\label{R}
\end{equation}
we emphasize that the number $n$ of Riemann invariants is allowed to 
be arbitrary. Solutions of this type, known as nonlinear interactions 
of $n$ planar
simple waves, were  discussed  in a series of publications 
\cite{Burnat1, Burnat3,  Perad2, Grundland}. Later, they
were  investigated by  Gibbons and Tsarev in  the context of the 
dispersionless KP hierarchy
\cite{Gibb81, Gibb94, GibTsa96, GibTsa99},  see also \cite{Ma}, and
the theory of  Egorov's integrable hydrodynamic chains \cite{Pavlov,
Pavlov1}.

We recall, see \cite{Tsarev}, that the requirement of  the
commutativity of the flows (\ref{R})
is equivalent to the following restrictions on their characteristic speeds:
\begin{equation}
\frac{\partial_j\lambda
^i}{\lambda^j-\lambda^i}=\frac{\partial_j\mu^i}{\mu^j-\mu^i}, ~~~
i\ne j,
~~~ \partial_j=\partial/\partial_{ R^j};
\label{comm}
\end{equation}
(no summation!)  Once these conditions are met, the general solution
of (\ref{R}) is given by the
implicit  `generalized hodograph'  formula \cite{Tsarev},
\begin{equation}
v^i(R)=y+\lambda^i(R)\ t+\mu^i(R) \ x, ~~~ i=1, ..., n,
\label{hod}
\end{equation}
where $v^i(R)$ are  characteristic speeds of the general flow
commuting with (\ref{R}), that is, the general solution of the linear
system
\begin{equation}
\frac{\partial_jv^i}{v^j-v^i}=\frac{\partial_j\lambda
^i}{\lambda^j-\lambda^i}=\frac{\partial_j\mu^i}{\mu^j-\mu^i}.
\label{comm1}
\end{equation}
Substituting ${\bf u}(R^1, ..., R^n)$ into (\ref{1}) and using
(\ref{R}), one readily arrives at the equations
\begin{equation}
(\lambda^i I_m+\mu^iA+B)\ \partial_i{\bf u}=0, ~~~~~ i=1, ..., n,
\label{2}
\end{equation}
implying that  $\lambda^i$ and $\mu^i$ satisfy the dispersion relation
\begin{equation}
{\rm det} (\lambda I_m+\mu A+ B)=0.
\label{dispersion}
\end{equation}
Thus, the construction of nonlinear interactions of $n$ planar simple
waves reduces to solving the equations  (\ref{comm}),
(\ref{2}) for ${\bf u}(R), \ \lambda^i(R), \ \mu^i(R)$ as functions
of the Riemann invariants $R^1, ..., R^n$. For $n\geq 3$ these
equations are highly overdetermined and
do not possess solutions in general. As demonstrated in \cite{Fer3}, 
the requirement of  the existence of nontrivial
3-component reductions is very  restrictive and implies,  in particular,  the
existence of  $n$-component reductions for arbitrary $n$.  We give 
the following
\vspace{0.5cm}

\noindent{\bf Definition.}
 {\it The system (\ref{1}) is said to be 
integrable if it possesses $n$-component reductions of
the form (\ref{R})  parametrized  by $n$ arbitrary functions of a
single argument}.
\vspace{0.5cm}

\noindent We refer to \cite{Fer3} for the motivation and supporting examples.

{\bf Remark 1.} In the case of linear systems ({\ref{1}), that is, in 
the case when both $A$ and $B$ are constant matrices, the equations 
(\ref{comm}) and (\ref{dispersion})
  imply $\lambda^i_j=\mu^i_j=0$, so that $\lambda^i=\lambda^i(R^i), \ 
\mu^i=\mu^i(R^i)$. Moreover, as follows from (\ref{2}),  $\partial_i 
u=\xi_i(R^i)$ where
$\xi_i(R^i)$ is the right eigenvector of the matrix $\lambda^i 
I_m+\mu^iA+B$. With the particular choice $\lambda^i=const, \ 
\mu^i=const, \ \xi_i=const$
the corresponding solutions represent the standard linear 
superposition of simple waves, $u=\sum f^i(x+\lambda^it+\mu^iy)\ 
\xi_i$.

In Sect. 2 we derive the integrability conditions for the two-component 
system (\ref{1}) assuming that the matrix $A$ is written in the 
diagonal form,
\begin{equation}
\left(
\begin{array}{c}
v \\
\ \\
w
\end{array}
\right)_t+
\left(
\begin{array}{cc}
a & 0 \\
\ \\
0 & b
\end{array}
\right)
\left(
\begin{array}{c}
v \\
\ \\
w
\end{array}
\right)_x+
\left(
\begin{array}{cc}
p&q \\
\ \\
r&s
\end{array}
\right)
\left(
\begin{array}{c}
v \\
\ \\
w
\end{array}
\right)_y=0;
\label{sys}
\end{equation}
such diagonalization is  always possible in the two-component 
situation. These  conditions constitute a complicated overdetermined
system (\ref{a}) - (\ref{qr}) of second order PDEs for  $a, \ b, \ p, 
\ q, \ r, \ s$ as functions of $v, w$, which  is in involution; a 
simple analysis shows that the class of integrable two-component 
systems is locally parametrised by 15 arbitrary constants.

{\bf Remark 2.} In principle,  the method described in Sect. 2 allows 
one to derive the integrability conditions in arbitrary coordinates, 
however, the formulasbecome extremely complicated. We were not able 
to find an invariant `tensor' formulation of the integrability 
conditions so far.

We prove  (Theorem 1 of Sect. 3) that an arbitrary two-component 
system (\ref{sys}) satisfying the integrability conditions possesses 
three conservation laws of hydrodynamic type and, thus, is 
symmetrizable in Godunov's sense \cite{Godunov}.

In Sect. 4 we demonstrate that all two-component integrable systems 
possess scalar pseudopotentials of the form
$$
\psi_t=f(\psi_y, \ v, \ w), ~~~ \psi_x=g(\psi_y, \ v, \ w).
$$
  According to the philosophy of \cite{Zakharov},
this indicates that (2+1)-dimensional integrable systems of 
hydrodynamic type can be obtained as dispersionless limits from the 
appropriate (2+1)-dimensional integrable soliton equations (possibly, 
nonlocal, differential-difference, etc). The corresponding 
pseudopotentials are quasiclassical limits of the associated linear 
Lax operators.
The construction of the `solitonic prototype' was sketched in the 
case when the dependence of $f$ and $g$ on $\psi_y$ is rational 
(trigonometric), leading to differential (difference) soliton 
equations.
We prove (Theorem 2 of Sect. 4) that the requirement of the existence 
of a scalar pseudopotential is, in fact, necessary and sufficient for 
the existence of the infinity of hydrodynamic reductions. This 
establishes the equivalence of the two approaches to  integrability 
of  (2+1)-dimensional hydrodynamic type systems. The quasi-classical 
$\overline\partial$-dressing approach to the solution of 
(2+1)-dimensional dispersionless systems  based on the 
pseudopotentials of the above type was proposed in the series of 
recent publications \cite{Kon1, Kon2, Kon3, Kon4, Bogdanov}. It is 
not completely clear at the moment how exact solutions describing 
nonlinear interactions of planar simple waves fit into this scheme.

Some explicit examples where the matrix $A$ is linearly degenerate 
(that is, $a_v=b_w=0$), are discussed in Sect. 5. These include a 
remarkable case where {\it both} matrices $A$ and $B$, as well as 
arbitrary linear combinations thereof,  are linearly degenerate.

We conclude this introduction by listing some known examples of 
two-component integrable systems written in the form (\ref{sys}).

{\bf Example 1.} Let us consider the system
$$
v_t+\frac{1}{v+w}\ v_x-\frac{1}{v+w} \ w_y=0, ~~~ w_t-\frac{1}{v+w}\ 
w_x+\frac{1}{v+w} \ v_y=0;
$$
here $a=\frac{1}{v+w}, \ b=-\frac{1}{v+w},$ etc. Introducing the 
variables $m=v+w, \ n=v-w$, one can rewrite these equations in the 
form
$$
(\partial_x+\partial_y) n+\frac{1}{2}\partial_t m^2=0, ~~~ \partial_t 
n+(\partial_x-\partial_y) \ln m=0
$$
leading, upon cross-differentiation, to the Boyer-Finley equation for 
$m^2=(v+w)^2$:
$$
\partial_t^2 m^2=(\partial_x^2-\partial_y^2)\ln m^2.
$$
The Boyer-Finley equation is known to be integrable, its hydrodynamic 
reductions were investigated, e.g., in \cite{Fer}.

{\bf Example 2.} A closely related example is
$$
v_t+\frac{1}{v+w}\ v_x+\frac{1}{v+w}\sqrt{\frac{v}{w}} \ w_y=0, ~~~ 
w_t-\frac{1}{v+w}\ w_x+\frac{1}{v+w}\sqrt{\frac{w}{v}} \ v_y=0;
$$
notice that the characteristic speeds $a$ and $b$ are the same as in 
the previous example! In the new variables $m=v-w, \ n=2\sqrt {vw}$, 
this system reduces to
$$
m_t+\frac{mm_x+nn_x}{m^2+n^2}+\frac{mn_y-nm_y}{m^2+n^2}=0, ~~~
n_t+\frac{nm_x-mn_x}{m^2+n^2}+\frac{mm_y+nn_y}{m^2+n^2}=0;
$$
in this form it appeared in out recent paper \cite{Fer3}. It was 
demonstrated, in particular, that the expression 
$\rho^2=m^2+n^2=(v+w)^2$ satisfies another version of the 
Boyer-Finley equation,
$$
\partial_t^2 \rho^2=(\partial_x^2+\partial_y^2)\ln \rho^2,
$$
corresponding to  different signature.

{\bf Example 3.} Here both matrices $A$ and $B$, as well as 
arbitrary linear combinations thereof, are linearly degenerate:
$$
v_t+wv_x+\frac{1}{w-v}(v_y+w_y)=0, ~~~ w_t+vw_x+\frac{1}{v-w}(v_y+w_y)=0.
$$
Introducing the variables $m=v+w, \ n=vw$, one can rewrite these equations as
$$
m_t+n_x=0, ~~~ n_t+mn_x-nm_x+m_y=0.
$$
This system was thoroughly investigated in \cite{Pavlov}, see also \cite{MaShabat, 
Shabat}.

{\bf Example 4.} We also looked at the integrable systems (\ref{sys}) 
whose chararacteristic speeds $a$ and $b$ are of the form 
$a=v+w+\epsilon v, \ b=v+w+\epsilon w, \ \epsilon={\rm const}$. The 
analysis showed that the only possible values for $\epsilon$ are 
$\epsilon =-1$ and $\epsilon =-2$. In the first case the matrix $A$ 
is linearly degenerate, see  Sect. 4 for 
the general form of the corresponding matrix $B$. In the case 
$\epsilon =-2$ we obtained the system
$$
v_t+(v-w)\ v_x+w_y=0, ~~~ w_t+(w-v)\ w_x+v_y=0,
$$
which is yet another first order form of the Boyer-Finley equation, 
indeed, this system reduces to that from the Example 1 after a simple 
change of variables $w\to -w, \ t\leftrightarrow x$.

\section{Derivation of the integrability conditions}

The integrability conditions  can be obtained as follows.
Looking for reductions of the system (\ref{sys}) in the form $v=v(R^1,
..., R^n)$, $w=w(R^1, ..., R^n)$ where the Riemann invariants satisfy the
equations
(\ref{R}), and substituting into (\ref{sys}), one  arrives at
$$
(\lambda^i+a \mu^i+p)\ \partial_iv+q\ \partial_iw=0,
~~~
r\ \partial_iv+(\lambda^i+b \mu^i+s)\ \partial_iw=0,
$$
(no summation!) so that $\lambda^i$ and $\mu^i$ satisfy the 
dispersion relation
$$
(\lambda^i+a \mu^i+p)(\lambda^i+b \mu^i+s)=qr.
$$
We assume that the dispersion relation defines an irreducible conic, 
that is, $a\ne b, \ r\ne 0, \ q\ne 0$. Notice that these conditions 
are equivalent to the requirement $rk [A, B]=2$. Setting 
$\partial_iv=\varphi^i\partial_iw$ one obtains the following
expressions for $\lambda^i$ and $\mu^i$ in terms of $\varphi^i$,
\begin{equation}
\displaystyle{\lambda^i=\frac{ar (\varphi^i)^2+(as-bp) 
\varphi^i-bq}{(b-a)\varphi^i}},  ~~~~
\displaystyle{\mu^i=\frac{r(\varphi^i)^2+(s-p)\varphi^i-q}{(a-b)\varphi^i}},
\label{lm}
\end{equation}
which define a rational parametrization of the dispersion relation.
The compatibility conditions of the equations
$\partial_iv=\varphi^i\partial_iw$ imply
\begin{equation}
\partial_i\partial_jw=\frac{\partial_j\varphi^i}{\varphi^j-\varphi^i}\partial_iw+\frac{\partial_i\varphi^j}{\varphi^i-\varphi^j}\partial_jw,
\label{3w}
\end{equation}
while the commutativity conditions (\ref{comm}) lead to the
expressions for $\partial_j \varphi^i, \ (i\ne j),$ in the form
\begin{equation}
\partial_j
\varphi^i=(...)\ \partial_jw;
\label{varphi}
\end{equation}
here dots denote a rational expression
in $\varphi^i, \varphi^j$
whose coefficients depend on $a, \ b, \ p, \ q, \ r, \ s$ and the 
first derivatives thereof.  We do not write them out explicitly due 
to their
complexity. To manipulate with these expressions we used symbolic
computations (Mathematica 5.0). Substituting the expressions for 
$\partial_j\varphi^i$ into (\ref{3w}) one obtains
\begin{equation}
\partial_i\partial_jw=(...) \ \partial_iw \partial_jw
\label{w}
\end{equation}
where, again, dots denote a rational expression in $\varphi^i, 
\varphi^j$. One can  see that  the  compatibility conditions of the 
equations (\ref{varphi}), that is,
$\partial_k\partial_j\varphi^i-\partial_j\partial_k\varphi^i=0$, are of
the form
$P\ \partial_jw\partial_kw=0$, where $P$ is a complicated rational
expression in $\varphi^i, \varphi^j, \varphi^k$ whose coefficients
depend on partial derivatives of $a, \ b, \ p, \ q, \ r, \ s$ up to 
second order (to obtain the integrability conditions it suffices to 
consider 3-component reductions setting $i=1, \ j=2, \ k=3$).
Requiring that  $P$ vanishes identically we obtain  the expressions 
for all second partial derivatives of the potentials $a, \ b, \ p, 
\ s$, as well as three relations among the second partial derivatives 
of $q$ and $r$. Similarly, the compatibility conditions
of the equations (\ref{w}),  that is, 
$\partial_k(\partial_i\partial_jw)-\partial_j(\partial_i\partial_kw)=0$, 
take the form $Q\ \partial_iw\partial_jw\partial_kw=0$ where, again, 
$Q$ is a rational
expression in $\varphi^i, \varphi^j, \varphi^k$. Equating $Q$ to zero 
one obtains (modulo  conditions obtained on the previous step) the 
expressions for  mixed partial derivatives  $q_{vw}$ and $r_{vw}$. 
The resulting set of the integrability conditions looks as follows.

\noindent {\bf Equations for $a$:}
\begin{eqnarray}
a_{vv}&=&\frac{\displaystyle{qa_vb_v+2qa_v^2+(s-p)a_va_w-ra_w^2}}{\displaystyle{(a-b)q}}+\frac{\displaystyle{a_vr_v}}{\displaystyle{r}}+\frac{\displaystyle{2a_vp_w-a_wp_v}}{\displaystyle{q}}, 
\nonumber \\
a_{vw}&=&a_v\frac{\displaystyle{a_w+b_w}}{\displaystyle{a-b}}+a_v(\frac{\displaystyle{q_w}}{q}+\frac{\displaystyle{r_w}}{r}), 
\label{a} \\
a_{ww}&=&\frac{\displaystyle{qa_vb_v+(s-p)a_vb_w+ra_w^2}}{\displaystyle{(a-b)r}}+\frac{\displaystyle{a_vs_w}}{r}+\frac{\displaystyle{a_wq_w}}{q}.
\nonumber
\end{eqnarray}
\\
\\

\noindent  {\bf Equations for $b$:}
\begin{eqnarray}
\nonumber \\
b_{vv}&=&\frac{\displaystyle{ra_wb_w+(p-s)a_vb_w+qb_v^2}}{\displaystyle{(b-a)q}}+\frac{\displaystyle{b_wp_v}}{q}+\frac{\displaystyle{b_vr_v}}{r}; 
\nonumber \\
b_{vw}&=&b_w\frac{\displaystyle{a_v+b_v}}{\displaystyle{b-a}}+b_w(\frac{\displaystyle{q_v}}{q}+\frac{\displaystyle{r_v}}{r}), 
\label{b} \\
b_{ww}&=&\frac{\displaystyle{ra_wb_w+2rb_w^2+(p-s)b_vb_w-qb_v^2}}{\displaystyle{(b-a)r}}+\frac{\displaystyle{b_wq_w}}{\displaystyle{q}}+\frac{\displaystyle{2b_ws_v-b_vs_w}}{\displaystyle{r}}.
\nonumber
\end{eqnarray}

\noindent  {\bf Equations for $p$:}
\begin{eqnarray}
p_{vv}&=&2\frac{r(a_vb_w-a_wb_v)+(s-p)a_vb_v}{(a-b)^2}+\frac{r_vp_v}{r}+\frac{p_vp_w}{q}+ 
\nonumber \\
&&\frac{\frac{r}{q} 
(2q_va_w-2a_vq_w+a_wp_w)-b_vp_v+2r_va_w-2a_v(s_v+p_v+r_w)+\frac{p-s}{q}(2p_va_w-a_vp_w)}{b-a}\nonumber 
\\
\ \nonumber \\
p_{vw}&=&2(s-p)\frac{a_vb_w}{(a-b)^2}-\frac{b_wp_v+(2s_w+p_w)a_v}{b-a}+{p_v}\left 
(\frac{q_w}{q}+\frac{r_w}{r}\right ), \label{p} \\
\  \nonumber \\
p_{ww}&=&2\frac{q(a_wb_v-a_vb_w)+(s-p)a_wb_w}{(a-b)^2}+\frac{(p-s)b_wp_v-qb_vp_v-2rs_wa_w-ra_wp_w}{(b-a)r}+ 
\nonumber \\
&& \frac{p_vs_w}{r}+\frac{q_wp_w}{q}. \nonumber
\end{eqnarray}

\noindent  {\bf Equations for $s$:}

\begin{eqnarray}
s_{vv}&=&2\frac{r(a_wb_v-a_vb_w)+(p-s)a_vb_v}{(a-b)^2}+\frac{(s-p)a_vs_w-ra_ws_w-2qp_vb_v-qb_vs_v}{(a-b)q}+ 
\nonumber \\
&&\frac{p_vs_w}{q}+\frac{r_vs_v}{r}, \nonumber \\
\  \nonumber \\
s_{vw}&=&2(p-s)\frac{a_vb_w}{(a-b)^2}-\frac{a_vs_w+(2p_v+s_v)b_w}{a-b}+{s_w}\left 
(\frac{q_v}{q}+\frac{r_v}{r}\right ), \label{s} \\
\  \nonumber \\
s_{ww}&=&2\frac{q(a_vb_w-a_wb_v)+(p-s)a_wb_w}{(a-b)^2}+\frac{q_ws_w}{q}+\frac{s_vs_w}{r} 
+ \nonumber \\
&&\frac{\frac{q}{r} 
(2r_wb_v-2b_wr_v+b_vs_v)-a_ws_w+2q_wb_v-2b_w(p_w+s_w+q_v)+\frac{s-p}{r}(2s_wb_v-b_ws_v)}{a-b}. 
\nonumber
\end{eqnarray}
\noindent {\bf Equations for  $q$ and $r$:}
\begin{eqnarray}
qr_{vv}+rq_{vv}&=&2(p-s)\frac{(p-s)a_wb_w+q(a_vb_w-a_wb_v)}{(a-b)^2} 
+q\frac{r_v}{r}\frac{qb_v+(s-p)b_w}{a-b}+   \nonumber \\
&&(s-p)\frac{2a_ws_w+2b_wp_w+b_wq_v}{a-b}+r\frac{(a_w-2b_w)q_w}{a-b}+ 
\nonumber \\
&&q\frac{a_wr_w+b_v(2p_w+2s_w+q_v)-2b_w(r_w+p_v+s_v)}{a-b}+ \nonumber \\
&&\frac{r}{q}q_w^2+\frac{q}{r}s_wr_v-q_wr_w+s_w(2p_w+q_v), \nonumber \\
\ \nonumber \\
q_{vw}&=&(s-p)\frac{qa_vb_v+(s-p)a_vb_w+ra_wb_w}{r(a-b)^2} 
+\frac{q_vq_w}{q}+\frac{p_vs_w}{r}+                     \nonumber \\
&& 
\frac{a_v(rq_w+qr_w)+(s-p)(a_vs_w+b_wp_v)+ra_ws_w+qp_vb_v}{r(a-b)}, 
\nonumber \\
\  \label{qr} \\
r_{vw}&=&(p-s)\frac{ra_wb_w+(p-s)a_vb_w+qa_vb_v}{q(a-b)^2} 
+\frac{r_vr_w}{r}+\frac{p_vs_w}{q}+                     \nonumber \\
&& 
\frac{b_w(rq_v+qr_v)+(p-s)(a_vs_w+b_wp_v)+ra_ws_w+qp_vb_v}{q(b-a)}, 
\nonumber \\
\ \nonumber \\
qr_{ww}+rq_{ww}&=&2(s-p)\frac{(s-p)a_vb_v+r(a_vb_w-a_wb_v)}{(a-b)^2} 
+r\frac{q_w}{q}\frac{ra_w+(p-s)a_v}{b-a}+   \nonumber \\
&&(p-s)\frac{2b_vp_v+2a_vs_v+a_vr_w}{b-a}+q\frac{(b_v-2a_v)r_v}{b-a}+ 
\nonumber \\
&&r\frac{b_vq_v+a_w(2s_v+2p_v+r_w)-2a_v(q_v+s_w+p_w)}{b-a}+ \nonumber \\
&&\frac{q}{r}r_v^2+\frac{r}{q}p_vq_w-r_vq_v+p_v(2s_v+r_w); \nonumber
\end{eqnarray}
notice that there are only two relations among the  second 
derivatives $q_{vv}, r_{vv}, q_{ww}, r_{ww}$.
These formulas are completely symmetric under the identification 
$v\leftrightarrow w, \ a \leftrightarrow b, \ p\leftrightarrow s, \ 
q\leftrightarrow r$. It can be verified that the equations (\ref{a}) 
- (\ref{qr}) are in involution and their general solution depends, 
modulo the coordinate transformations $v= \varphi (\tilde v), \ w= 
\psi (\tilde w)$, on 15 arbitrary constants. Thus, we have 
established the existence of a 15-parameter family of  integrable 
systems of the form (\ref{sys}).

Once the integrability conditions (\ref{a}) - (\ref{qr}) are 
satisfied,  the general solution of the involutive system 
(\ref{varphi}), (\ref{w}) for $\varphi^i$ and $w$  will depend on 
$2n$ arbitrary functions of a single argument (indeed, one can 
formulate the Goursat problem for this system specifying $\varphi^i$ 
along the $R^i$-coordinate line and specifying the restriction of $w$ 
to each of the coordinate lines). This has to be considered up to 
reparametrizations of the form $R^i\to f^i(R^i)$. Thus, the general 
$n$-component reduction  depends on $n$ essential functions of a 
single argument.
This justifies the definition of the integrability given in the 
Introduction. The system (\ref{varphi}), (\ref{w}) governing 
$n$-component reductions will be called the generalized 
Gibbons-Tsarev system
(it was derived by Gibbons and Tsarev  \cite{GibTsa96} in the context 
of the dispersionless KP equation).

{\bf Remark.} Rewriting the equations $(\ref{a})_2$ and $(\ref{b})_2$ 
in the form $d\ln (qr)=\Omega$, where
$$
\Omega=\left(\frac{b_{vw}}{b_w}+\frac{a_v+b_v}{a-b}  \right)\ 
dv+\left(\frac{a_{vw}}{a_v}+\frac{a_w+b_w}{b-a}  \right)\ dw,
$$
(we assume $a_v\ne 0, \ b_w\ne 0$), one obtains the condition 
$d\Omega=0$ which involves the matrix $A$ only. Obviously, the same 
condition holds for an arbitrary matrix in the linear pencil
$\alpha A + \beta B$ (written in the diagonal form). The object 
$d\Omega$ first appeared in \cite{Ferrec1, Ferrec2} as one of the 
basic reciprocal invariants of two-component hydrodynamic type 
systems.

\section{Conservation laws}

In this section we prove the following

\begin{theorem}
Any two-component  (2+1)-dimensional system of hydrodynamic type 
which passes the integrability test necessarily possesses three 
conservation laws of hydrodynamic type and, hence,
is symmetrizable in Godunov's sense \cite{Godunov}.
\end{theorem}

This explains the observation made in our recent publication \cite{Fer3}.
To obtain the proof we first transform the system into the form (\ref{sys}).
Looking for conservation laws in the form
$$
h(v, w)_t+g(v, w)_x+f(v, w)_y=0,
$$
one readily obtains
$$
g_v=a\ h_v, ~~~~ g_w=b\ h_w
$$
and
$$
f_v=p\ h_v+r\ h_w, ~~~~ f_w=q\ h_v+s\ h_w.
$$
The consistency condition $g_{vw}=g_{wv}$ implies
\begin{equation}
h_{vw}=\frac{a_w}{b-a}h_v+\frac{b_v}{a-b}h_w,
\label{vw}
\end{equation}
while the consistency condition $f_{vw}=f_{wv}$ results in
$$
p_wh_v+p\left(\frac{a_w}{b-a}h_v+\frac{b_v}{a-b}h_w\right)+r_wh_w+rh_{ww}=
$$
$$
s_vh_w+s\left(\frac{a_w}{b-a}h_v+\frac{b_v}{a-b}h_w\right)+q_vh_v+qh_{vv}.
$$
The last formula can be rewritten in the form
\begin{equation}
\begin{array}{c}
h_{vv}=\frac{1}{q}\left(\frac{s-p}{a-b}a_w+p_w-q_v\right)\ h_v+\frac{l}{q}, \\
\ \\
h_{ww}=\frac{1}{r}\left(\frac{s-p}{a-b}b_v+s_v-r_w\right)\ h_w+\frac{l}{r},
\end{array}
\label{vvww}
\end{equation}
where the equations for the auxiliary variable $l$ can be obtained 
from the compatibility conditions
$(h_{vv})_w=(h_{vw})_v$ and $(h_{ww})_v=(h_{vw})_w$:
\begin{equation}
\begin{array}{c}
l_{v}=\left(\frac{r_v}{r}+\frac{b_v}{a-b}\right)\ l-
\frac{2b_vp_v(b-a)+2a_vb_v(p-s)+4ra_wb_v-ra_vb_w}{(a-b)^2} \ h_w- \\
\ \\
\frac{a_vb_vq+a_wb_wr+(a-b)ra_w\frac{q_w}{q}+(a_wb_v-a_vb_w)(p-s)+(b-a)a_w(s_v-r_w) 
+(a-b)a_vs_w-ra_w^2 }{(a-b)^2}\ h_v,
\ \\
\ \\
l_{w}=\left(\frac{q_w}{q}+\frac{a_w}{b-a}\right)\ l-
\frac{2a_ws_w(a-b)+2a_wb_w(s-p)+4qa_wb_v-qa_vb_w}{(a-b)^2} \ h_v- \\
\ \\
\frac{a_vb_vq+a_wb_wr+(b-a)qb_v\frac{r_v}{r}+(a_wb_v-a_vb_w)(s-p)+(a-b)b_v(p_w-q_v) 
+(b-a)b_wp_v-qb_v^2 }{(a-b)^2}\ h_w.
\end{array}
\label{l}
\end{equation}
One can verify that the compatibility conditions $l_{vw}=l_{wv}$ are 
satisfied identically by virtue of (\ref{a})-(\ref{qr}). Thus, the 
system of equations (\ref{vw}), (\ref{vvww}) and (\ref{l})
for conservation laws  is in  involution and its solution space is 
three-dimensional.

\section{Pseudopotentials}

In this section we prove that any integrable system (\ref{sys}) 
possesses a scalar pseudopotential depending, in some cases, on the 
auxiliary parameter $\lambda$. We begin with some supporting examples.

{\bf Example 5.} The linearly degenerate system
$$
m_t+n_x=0, ~~~ n_t+mn_x-nm_x+m_y=0
$$
from the Example 3 possesses the pseudopotential
$$
\psi_t=-(\lambda+m)\psi_x, ~~~  \psi_y=(\lambda^2+\lambda m+n)\psi_x;
$$
we emphasize that the parameter $\lambda$ is essential here, allowing 
one to recover the full system for $m, n$ from the consistency 
condition $\psi_{ty}=\psi_{yt}$.

{\bf Example 6.} The dispersionless KP equation, 
$(u_t-uu_x)_x=u_{yy}$, rewritten in the two-component form
$$
u_y=w_x, ~~~ w_y=u_t-uu_x,
$$
  possesses the pseudopotential
$$
\psi_t=\frac{1}{3}\psi_x^3+u\psi_x+ w, ~~~ \psi_y=\frac{1}{2}\psi_x^2+ u,
$$
see \cite{Zakharov}.

{\bf Example 7.}  The Boyer-Finley equation, $u_{tt}=(\ln u)_{xy}$, 
rewritten in the 2-component form
$$
u_t=w_y, ~~~ w_t=u_x/u,
$$
possesses the pseudopotential
$$
\psi_t=\ln u-\ln \psi_y, ~~~ \psi_x=w-\frac{u}{  \psi_y}.
$$

\noindent Further examples of integrable (2+1)-dimensional equations 
possessing pseudopotentials of the above type can be found in 
\cite{Zakharov, Pavlov1, Kon1}. It is a remarkable fact that in all 
examples constructed in \cite{Pavlov1} the existence of such 
pseudopotentials manifests the equivalence of the corresponding 
(2+1)-dimensional system to a pair of  commuting (1+1)-dimensional 
hydrodynamic chains.

\bigskip

In the general case of system (\ref{sys}) we look for a 
pseudopotential in the form
\begin{equation}
\psi_t=f(\psi_y, \ v, \ w), ~~~ \psi_x=g(\psi_y, \ v, \ w).
\label{Lax}
\end{equation}
Writing out the consistency condition $\psi_{tx}=\psi_{xt}$, 
expressing $v_t, \ w_t$  by virtue of (\ref{sys}) and equating to 
zero coefficients at $v_x, v_y, w_x, w_y$, one arrives at the 
following expressions for the first derivatives $f_v, f_w, f_{\xi}$ 
and $g_{\xi}$ (we adopt the notation $\xi \equiv \psi_y$):
\begin{equation}
\begin{array}{c}
f_v=-a\ g_v, ~~~ f_w=-b\ g_w, \\
\ \\
f_{\xi}=\frac{{b\left(p+r\frac{g_w}{g_v}\right)-a\left(s+q\frac{g_v}{g_w}\right)}}{{a-b}},
\end{array}
\label{f}
\end{equation}
and
\begin{equation}
  g_{\xi}=\frac{{s+q\frac{g_v}{g_w}-p-r\frac{g_w}{g_v}}}{{a-b}}.
\label{g}
\end{equation}
The consistency conditions of the equations (\ref{f}) imply the 
following expressions for the second partial derivatives $g_{vw}, 
g_{vv}, g_{ww}$:
\begin{equation}
\begin{array}{c}
g_{vw}=\frac{a_w}{b-a}\ g_v+\frac{b_v}{a-b}\ g_w, \\
\ \\
g_{vv}=\frac{g_v[g_w^2(r(b_v-a_v)+(a-b)r_v)+g_vg_w((a-b)p_v+(s-p)a_v-ra_w)+qa_vg_v^2]}{(a-b) 
r g_w^2}, \\
\  \\
g_{ww}=\frac{g_w[g_v^2(q(a_w-b_w)+(b-a)q_w)+g_vg_w((b-a)s_w+(p-s)b_w-qb_v)+rb_wg_w^2]}{(b-a) 
q g_v^2}.
\end{array}
\label{g2}
\end{equation}
The compatibility conditions of the equations (\ref{g}), (\ref{g2}) 
for $g$, namely, the conditions  $g_{\xi vv}=g_{vv \xi}, \ g_{\xi 
vw}=g_{vw \xi} $, etc., are of the form
$P(g_v, g_w)=0$, where $P$ denotes a rational expression in $g_v, \ 
g_w$ whose coefficients are functions of $a, b, p, q, r, s$ and their 
partial derivatives  up to the second order. Equating all these 
expressions to zero (they are required to be zero identically in 
$g_v, g_w$), one obtains the  set of conditions which are necessary and 
sufficient for the existence of pseudopotentials of the form 
(\ref{Lax}).
It is a truly remarkable fact that these conditions {\it identically} 
coincide with the integrability conditions (\ref{a}) - (\ref{qr}).
Thus, any  system satisfying the integrability conditions  (\ref{a}) 
- (\ref{qr}) possesses pseudopotentials of the form (\ref{Lax}) 
parametrized by 4 arbitrary integration constants, indeed, one can 
arbitrarily prescribe
the values of $g, \ g_v, \ g_w$ and $f$ at any initial point; the 
rest is completely determined by the involutive system (\ref{g}), 
(\ref{g2}) and (\ref{f}). Notice, however, that the transformation
$\psi \to \lambda \psi+ \mu x+ \nu y+\eta t$ allows one to eliminate 
all these constants in the general situation (see Example 5 where one 
of these constant survives and is essential).

We have established the following

\begin{theorem}
The class of two-component (2+1)-dimensional systems of hydrodynamic 
type possessing infinitely many hydrodynamic reductions coincides 
with the class of systems possessing a scalar pseudopotential of the 
form (\ref{Lax}).
\end{theorem}

{\bf Remark. } The pseudopotential (\ref{Lax}) readily implies a 
pseudopotential for the corresponding generalized Gibbons-Tsarev 
system (\ref{varphi}), (\ref{w}). Indeed, differentiating
the equations (\ref{Lax}) by $y$ and introducing $\xi=\psi_y$, one obtains
$$
\xi_t=\partial_y f(\xi, \ v, \ w)=f_{\xi} \xi_y+f_v  v_y+f_w w_y, 
~~~ \xi_x=\partial_yg(\xi, \ v, \ w)=g_{\xi} \xi_y+g_v v_y+g_w w_y.
$$
Assuming now that $\xi, v, w$ are functions of $n$ Riemann invariants 
$R^1, ..., R^n$ which satisfy the equations (\ref{R}), one arrives at
$$
\xi_i  \lambda^i=f_{\xi} \xi_i+f_v  v_i+f_w w_i, ~~~ \xi_i 
\mu^i=g_{\xi} \xi_i+g_v v_i+g_w w_i.
$$
Substituting here $v_i=\varphi^i w_i$, the expressions (\ref{lm}) for 
$\lambda^i$ and $\mu^i$ in terms of $\varphi^i$  (see Sect. 2),  and 
taking into account the formulae (\ref{f}), (\ref{g}), one ends up 
with
\begin{equation}
\xi_i=\frac{(a-b)\varphi^i}{r \varphi^i/g_v-q/g_w}\ w_i.
\label{xi}
\end{equation}
Equations (\ref{xi}) define a scalar pseudopotential for the 
generalized Gibbons-Tsarev system (\ref{varphi}), (\ref{w}), that is, 
the consistency conditions of (\ref{xi}) imply the equations
(\ref{varphi}), (\ref{w}).

\section{Examples}

The equations (\ref{a}) - (\ref{qr}) are particularly convenient to 
analyze when the matrix $A$ is {\it given} (we emphasize that $a$ and 
$b$ cannot be arbitrary). The corresponding matrix $B$ is then 
defined up to a natural equivalence
$$
B\to \mu B+\nu A+\eta I_2
$$
generated by a linear change of the independent variables  in the 
equations (\ref{sys}): $\tilde t = t, \ \tilde x=x, \ \tilde y=\mu 
y+\nu x+\eta t$; here $\mu, \nu, \eta$ are arbitrary constants. 
Moreover, one has a freedom
of the coordinate transformations $v= \varphi (\tilde v), \ w= \psi 
(\tilde w)$ preserving the diagonal form of $A$. These 
transformations do not change $a, b, p, s$ and transform $q$ and $r$ 
according to the formulas
  $$
\tilde q=q\frac{\psi ' (\tilde w)}{\varphi '(\tilde v)}, ~~~ \tilde 
r=r\frac{\varphi '(\tilde v)}{\psi ' (\tilde w)}.
$$
The classification results presented below are carried out up to this 
natural equivalence.

In this section we  concentrate on the case when the matrix $A$ is 
linearly degenerate,
that is, $a_v=b_w=0$. There are three essentially different cases to consider:
$$
{\rm Case~1:} ~~
  A=\left(
\begin{array}{cc}
w & 0 \\
\ \\
0 & v
\end{array}
\right); ~~~
{\rm Case~2:} ~~
  A=\left(
\begin{array}{cc}
\alpha & 0 \\
\ \\
0 & \beta
\end{array}
\right); ~~~
{\rm Case~3:} ~~
  A=\left(
\begin{array}{cc}
w & 0 \\
\ \\
0 & \beta
\end{array}
\right);
$$
here $\alpha$ and $\beta$ are arbitrary constants. Notice that 
without any loss of  generality one can set $\alpha =1, \ 
\beta=0$. Below we restrict ourselves to the symmetric cases 1 and 2, 
and show that there is a multi-parameter freedom in the formulas for 
$B$.

\noindent {\bf Case 1.} Substituting $a=w, \ b=v$ into the 
integrability conditions (\ref{a}) - (\ref{qr}), one obtains the overdetermined 
system for $p, q, r, s$ which can be explicitly integrated (the 
integration is fairly straightforward so that we skip the details).
Up to the  equivalence mentioned above we have
$$
p=\frac{f(w)}{w-v}-\alpha w^2, ~~~ q= \frac{f(v)}{w-v}, ~~~ 
r=\frac{f(w)}{v-w}, ~~~ s=\frac{f(v)}{v-w}-\alpha v^2,
$$
where $f$ is a cubic polynomial, $f(z)=\alpha z^3+\beta z^2 +\gamma z 
+\delta$, and $\alpha, \beta, \gamma, \delta$ are arbitrary 
constants. A remarkable property of this example is that
{\it any} matrix in the linear pencil $B+\mu A$ is also linearly 
degenerate. In the particular case $\alpha=\beta=\gamma=0, \  \delta 
=1$ one has
$$
v_t+wv_x+\frac{1}{w-v}(v_y+w_y)=0, ~~~ w_t+vw_x+\frac{1}{v-w}(v_y+w_y)=0.
$$
This system possesses three conservation laws
$$
(v+w)_t+(vw)_x=0,
$$
$$
(v^2+vw+w^2)_t+(vw(v+w))_x-(v+w)_y=0
$$
and
$$
(v^3+v^2w+vw^2+w^3)_t+(vw(v^2+vw+w^2))_x-(v+w)^2_y=0.
$$
  Introducing the variables $m=v+w, \ n=vw$, one can rewrite this system as
$$
m_t+n_x=0, ~~~ n_t+mn_x-nm_x+m_y=0.
$$
In this form it was thoroughly investigated in \cite{Pavlov}, see 
also \cite{MaShabat}.

\noindent {\bf Case 2.} Here $a$ and $b$ are constants, $a\ne b$. The 
corresponding equations for $p, q, r, s$ take the form
$$
p_{vv}=p_v\left(\frac{p_w}{q}+\frac{r_v}{r}\right), ~~~ 
p_{vw}=p_v\left(\frac{q_w}{q}+\frac{r_w}{r}\right), ~~~ 
p_{ww}=\frac{p_wq_w}{q}+\frac{s_wp_v}{r},
$$
$$
s_{vv}=\frac{s_vr_v}{r}+\frac{s_wp_v}{q}, ~~~ 
s_{vw}=s_w\left(\frac{q_v}{q}+\frac{r_v}{r}\right), ~~~ 
s_{ww}=s_w\left(\frac{s_v}{r}+\frac{q_w}{q}\right),
$$
$$
qr_{vv}+rq_{vv}=\frac{qr_v^2}{r}-q_vr_v+p_v(2s_v+r_w)+\frac{rq_wp_v}{q},
$$
$$
qr_{ww}+rq_{ww}=\frac{rq_w^2}{q}-q_wr_w+s_w(2p_w+q_v)+\frac{qr_vs_w}{r},
$$
$$
q_{vw}=\frac{q_vq_w}{q}+\frac{p_vs_w}{r}, ~~~ 
r_{vw}=\frac{r_vr_w}{r}+\frac{p_vs_w}{q}.
$$
These equations imply, in particular, that $(p_v/qr)_w=0, \ 
(s_w/qr)_v=0$ so that, after the appropriate reparametrization $v\to 
f(v), \ w \to g(w)$, one can set
$p_v=s_w=qr$ (provided $p_v\ne 0, \ s_w \ne 0$). With this 
simplification, the above equations reduce to
$$
p_v=qr, ~~ p_w=q_v, ~~~~~~~~ s_v=r_w, ~~ s_w=qr,
$$
along with the following overdetermined system for $q$ and $r$:
\begin{equation}
\begin{array}{c}
q_{vv}=(qr)_w, ~~~ q_{vw}=\frac{q_vq_w}{q}+q^2r, ~~~ 
q_{ww}=\frac{q_w^2}{q}+2qq_v-\frac{q_wr_w}{r}, \\
\ \\
r_{ww}=(qr)_v, ~~~ r_{vw}=\frac{r_vr_w}{r}+qr^2, ~~~ 
r_{vv}=\frac{r_v^2}{r}+2rr_w-\frac{q_vr_v}{q}.
\end{array}
\label{fin}
\end{equation}
This system  is in involution with the general solution depending on 
6 arbitrary constants.
Equations for $q_{vw}$ and $r_{vw}$ yield the 
Liouville equation for $\ln (q r)$ and the linear wave equation for 
$\ln (q/r)$, implying the following functional ansatz for these 
variables:
\begin{equation}
q = \frac{f'(v)^{1/2} g'(w)^{1/2}}{f(v) + 
g(w)} \frac{m(f(v))}{n(g(w))}, \quad
r = \frac{f'(v)^{1/2} 
g'(w)^{1/2}}{f(v) + g(w)} 
\frac{n(g(w))}{m(f(v))}.
\label{qr1}
\end{equation}
Setting
\begin{equation}
(f')^{3/2} 
= P(f), \quad (g')^{3/2} = Q(g),
\end{equation}
and substituting 
(\ref{qr1}) into the remaining equations  (\ref{fin}), we obtain the 
following four functional-differential equations for $P(f), Q(g), 
m(f), n(g)$:
\begin{eqnarray*}
&&\hspace{2.5cm} \left [P'' (f+g)^2 - 
4 P' (f+g) + 6 P \right ] m + \\
&&\left [4 P' (f+g)^2 - 6 P (f+g) 
\right ] m' + 3 P (f+g)^2 m'' =
\left [ 2 Q' (f+g) - 6 Q \right ] n, 
\\
\ \\
&&\hspace{2.5cm}\left [Q'' (f+g)^2 - 4 Q' (f+g) + 6 Q \right ] n + 
\\
&&\left [4 Q' (f+g)^2 - 6 Q (f+g) \right ] n' + 3 Q (f+g)^2 n'' 
=
\left [ 2 P' (f+g) - 6 P \right ] m, \\
\ \\
 &&\hspace{2cm}P m' + Q' 
(f+g) n' + Q \left [ (f+g) n'' - n' \right ] = 0,  \\
\ \\
&&\hspace{2cm}Q 
n' + P' (f+g) m' + P \left [ (f+g) m'' - m' \right ] = 
0.
\end{eqnarray*}
These equations yield 
$$
P(f) = \frac{\alpha 
f^3 + \beta f^2 + \gamma f + \delta}{m(f)}, \quad 
Q(g) = 
\frac{\alpha g^3 - \beta g^2 + \gamma g - \delta}{n(g)},
$$
where 

$$
(\ln m)' = \frac{A f}{\alpha f^3 + \beta f^2 + \gamma f + 
\delta}, \quad 
(\ln n)' = - \frac{A g}{\alpha g^3 - \beta g^2 + 
\gamma g - \delta}.
$$
Here, $\alpha, \beta, \gamma, \delta$ and $A$ 
are arbitrary constants.
If $A=0$ and $m=n=1$, then both $P$ and $Q$ 
are cubic polynomials in $f$ and $g$, implying that equations for $f$ 
and $g$ can be solved in terms  of elliptic functions (this case was considered in \cite{Fer3}).

\section{Conclusion}

In this paper 
we gave the characterization of two-component (2+1)-dimensional 
integrable systems of hydrodynamic type, showing 
that
\begin{itemize}
\item[\sf --] there exists a 15-parameter 
family of such systems;
\item[\sf --] all integrable systems are 
symmetrizable in Godunov's sense;
\item[\sf --] the system is 
integrable iff it possesses a scalar pseudopotential.
\end{itemize}
We 
have also constructed nontrivial explicit examples of 
integrable two-component  (2+1)-dimensional systems of hydrodynamic 
type for which one of the matrices of the system is linearly 
degenerate. 

The important problem remaining is to clarify the differential geometry of the full set of integrability conditions 
 (\ref{a}) - (\ref{qr}) expressing them in invariant form in terms of the corresponding matrices $A$ and $B$.

\section*{Acknowledgements}

We are grateful to   M. Pavlov and V. Sokolov
  for stimulating discussions.

\end{document}